# Superinjection in Diamond P-I-N Diodes: Bright Single-Photon Electroluminescence of Color Centers Beyond the Doping Limit

Igor A. Khramtsov and Dmitry Yu. Fedyanin[*]

*Laboratory of Nanooptics and Plasmonics, Moscow Institute of Physics and Technology, 141700 Dolgoprudny, Russian Federation*

[*]email: dmitry.fedyanin@phystech.edu

**ABSTRACT**

Efficient generation of single photons on demand at a high repetition rate is a key to the practical realization of quantum-communication networks and optical quantum computations. Color centers in diamond and related wide-bandgap semiconductors are considered to be the most promising candidates for building such single-photon sources owing to the outstanding emission properties at room temperature. However, efficient electrical excitation of color centers in most materials remains a challenge due to the inability to create a high density of free carriers. Here, we predict a superinjection effect in diamond p-i-n diodes. By employing a comprehensive theoretical approach, we numerically demonstrate that one can overcome the doping problem in diamond and inject four orders of magnitude more electrons into the i-region of the diamond p-i-n diode than the doping of the n-region allows. This high density of free electrons can be efficiently used to boost the single-photon electroluminescence process and enhance the brightness of the diamond single-photon source by more than three orders of magnitude. Moreover, we show that such a high single-photon emission rate can be achieved at exceptionally low injection current densities of only 0.001 A/mm$^2$, which creates the backbone for the development of low-power and cost-efficient diamond quantum optoelectronic devices for quantum information technologies.

# I. INTRODUCTION

Single color centers in wide-bandgap semiconductors are currently considered as promising candidates for a practical single-photon source – the device which is a key element for most applications of quantum information science [1–3]. Among several materials, such as silicon carbide [4], zinc oxide [5], gallium nitride [6], carbon nanotubes [7], and 2D semiconductors [8,9], that can host color center, diamond is one of the most attractive candidates. Its greatest advantage is an exceptionally weak electron-phonon interaction [10]. Diamond features a record Debye temperature of 2200 K [11]. Therefore, the emission spectra of many point defects in the crystal lattice of diamond (such as the silicon-vacancy (SiV) [12–14] and germanium-vacancy (GeV) centers [12,15]) exhibit a remarkably sharp peak with a dim phonon-sideband emission even at room temperature [12]. For example, more than 75% of photons emitted by the SiV center go into the zero-phonon line (ZPL) [13,14]. At the same time, the spectral width of the ZPL is less than 1 nm [13]. These defects can be created with high positioning accuracy (30-50 nm [16]) at any point of the nanostructure using focused ion beam implantation. In addition, due to the large bandgap, diamond is transparent in the visible and in infrared, while the indirect nature of the bandgap ensures a very low background luminescence level under both optical and electrical excitations. Such remarkable properties of color centers in diamond make them extremely attractive for room temperature applications [7–9,17–19].

The outstanding optical properties combined with the possibility to trigger color centers electrically on demand [20–25] are exactly what is required for building practical quantum information devices which operate at room temperature. However, it is not easy to efficiently excite color centers in diamond electrically by embedding them in a p-n or p-i-n diode [20–24]. Diamond is a material at the interface between semiconductors and insulators. This means that the conductivity of diamond cannot be as high as that of conventional semiconductors, such as silicon or gallium arsenide. The problem is especially pronounced in n-type samples since the activation energy of donors in diamond (~0.6 eV [26]) is an order of magnitude higher than in silicon (0.03–0.07 eV [27]). This high activation energy along with the unavoidable acceptor-type defects, which compensate donors in n-type samples, reduce the maximum density of free electrons at room temperature to ~$10^{10}$ cm$^{-3}$ [26,28] (see Fig. 1(a)). This is much lower than what can be created in most semiconductor materials. In turn, it was recently shown that the photon emission rate by color centers is determined by the electron and hole capture processes [29,30]. It is proportional to the densities of free electrons and holes in the vicinity of the color center, respectively. Therefore, due to the high activation energy of donors in diamond, the maximum photon emission rate is limited by approximately $\Phi \times \min(c_p p_{eqp}, c_n n_{eqn})$, where $n_{eqn}$ is the density of free electrons in the n-type region of the diamond structure in equilibrium, $p_{eqp}$ is the density of holes in the p-type region in equilibrium, $\Phi$ is the quantum efficiency of the color center, $c_n$ and $c_p$ are the capture rate constants, which characterize the processes of the electron and hole capture by the color center (for details, see [29–31]). Simple estimations show that at room temperature, the photon emission rate should not exceed ~1 kcps for most diamond samples [Fig. 1(b)]. Taking into account that typically only a few percent of photons can be collected by an optical objective [2], the observed photon count rate is nearly two orders of magnitude lower, which is not enough for practical applications. At high temperatures, electrically pumped color centers can be significantly brighter [29], but the emission properties of color centers are inferior to those at room temperature and are less suited for quantum optics applications.

Here we demonstrate that using the self-gating effect in a diamond p-i-n diode, even at room temperature, the single-photon electroluminescence (SPEL) rate of color centers embedded in a p-i-n diode can be increased to about 1 Mcps, which is three orders of magnitude above the doping limit. Moreover, we show that this rate can be achieved at current densities as low as 0.001 A/mm$^2$, which is particularly favorable for practical applications.

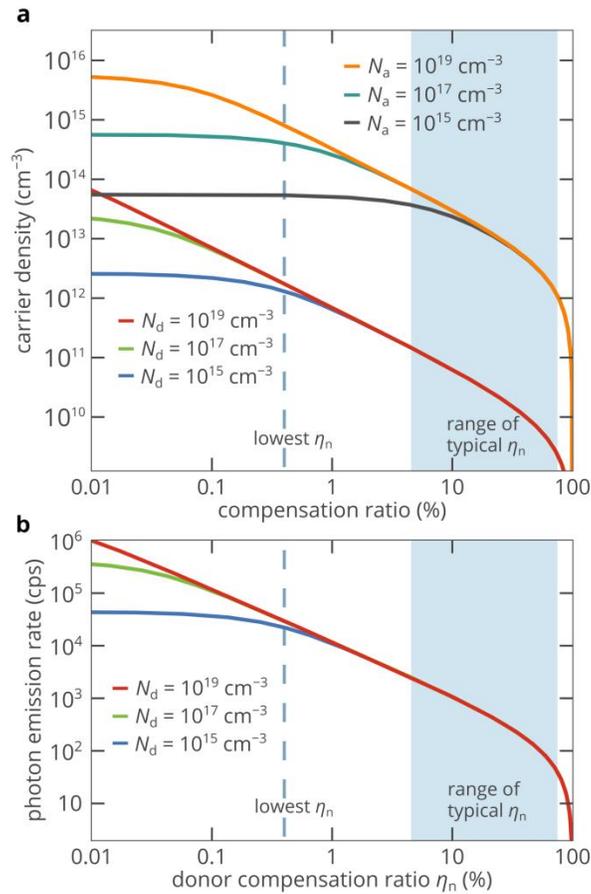

FIG. 1. (a) Free electron density in n-type diamond versus donor compensation ratio $\eta_n$. Also plotted is the dependence of the density of holes in p-type diamond as a function of the acceptor compensation ratio $\eta_p$. The blue-shaded area shows the range of donor compensation ratios

which are typical for n-type diamond samples, while the blue dashed vertical line indicates the lowest reported, to the best of our knowledge, donor compensation ratio of $\eta_n = 0.4\%$ [32]. (b) Estimated maximum photon emission rate of the color center as a function of the donor compensation ratio. The density of holes in the vicinity of the color center is assumed to be equal to $3.2\times10^{14}$ cm$^{-3}$, and the quantum efficiency is set to 100%.

## II. RESULTS

Figure 2(a) shows a schematic illustration of the single-photon emitting diamond p-i-n diode. The color center is incorporated in the i-region of the diode to reduce its interaction with impurities and defects in the n-type and p-type regions. Due to the high activation energies of donors and acceptors, such a diamond diode cannot be described analytically, especially at high forward bias voltages. Therefore, we performed self-consistent numerical simulations of the electron and hole transport based on the Poisson equation, drift-diffusion current equations, and the electron and hole continuity equations using the nextnano++ software (nextnano GmbH, Munich, Germany). The p-region is doped with boron at a concentration of $10^{18}$ cm$^{-3}$. The boron acceptors are partially compensated by donor-type defects, the compensation ratio is $\eta_p = 1\%$ [33], which corresponds to the density of holes of $3.2\times10^{14}$ cm$^{-3}$ [Fig. 1(a)]. The concentration of phosphorus donors in the n-region is also equal to $10^{18}$ cm$^{-3}$, but the donor compensation ratio is significantly higher and equals $\eta_n = 10\%$, which is typical for n-type diamond samples [26,28,34]. For other parameters used in the simulations, see Table S1 in Supplemental Material [35]. In equilibrium, the maximum density of electrons in the p-i-n diode is found in the n-region and equals $n_{eqn} = 6.3\times10^{10}$ cm$^{-3}$ [Fig. 1(a)], which virtually limits the photon emission rate of the color center to about $c_n n_{eqn} = 1000$ s$^{-1}$ at low and moderate injection currents. However, our simulations show that at high injection levels, i.e., at high forward bias voltages, the situation is significantly different.

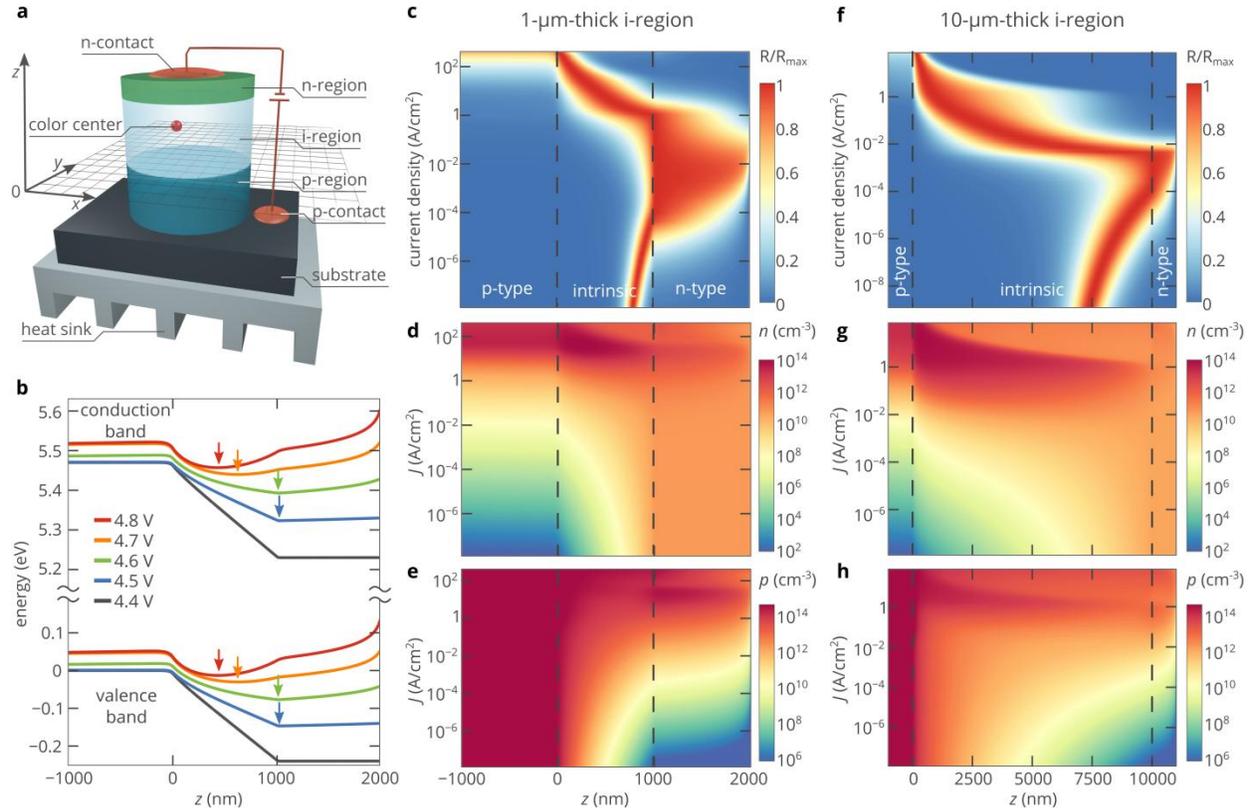

FIG. 2. (a) Schematic illustration of the single-photon emitting diamond p-i-n diode. The substrate is mounted to a heat sink to maintain the diode temperature at about 300 K. (b) Energy band diagram for the p-i-n diode with a 1-μm-thick i-region. The arrows next to the curves indicate the bottoms of the potential wells for electrons if such wells exist. (c) Distribution of the normalized SPEL rate as a function of the pump current. At each current, the SPEL rate $R(z)$ is normalized to the maximum achievable SPEL rate at this pump current $R_{max}$. The distribution of the non-normalized SPEL rate is shown in Fig. S1(a) in Supplemental Material [35]. (d,e) Evolution of the electron (d) and hole (e) distributions in the p-i-n diode with the pump current. In panels c-e, the thickness of the i-region of the p-i-n diode is equal to 1 μm. (f-

h) Spatial distributions of normalized SPEL rate (f), electron density (g) and hole density (h) as a function of the pump current. In panels (f)-(h), the thickness of the i-region is equal to 10 μm. The non-normalized SPEL rate for the diode with a 10-μm-thick i-region is shown in Fig. S1(b) in Supplemental Material [35].

Figure 2(c) shows normalized photon emission rate of the color center as a function of the injection current and the position of the color center in the p-i-n diode with a 1-μm-thick intrinsic region, which is obtained using the results of the numerical simulations of the electron and hole transport in the diamond diode and the recently developed theoretical model for electroluminescence of SiV centers. This model is discussed in detail in Refs. [29,31]. It is important to note here that Fig. 2 and most other figures discussed below are valid for most color centers in diamond which have two charge states: negatively charged and neutral. The reason for this is that the process of color-center electroluminescence is essentially based on the electron and hole exchange between the color center and the diamond crystal. This implies the processes of free-carrier capture (both electrons and holes) and free-exciton capture [29–31,36]. At room and higher temperatures, the contribution of the latter process is usually significantly weaker than that of the former (for details, see Ref. [30]). The hole capture cross-section by the negatively charged color center is mostly determined by the properties of diamond rather than by the internal structure of the defect [29,31], while the electron capture cross-section by the color center in the neutral charge state is roughly equal to the lattice constant of diamond and can only slightly vary from one defect to another [37,38], which is due to the different structure of the defects. Returning to Fig. 2(c), one can easily distinguish three injection regimes: the first is at current densities below $10^{-5}$ A/cm$^2$, the second is in the range from $10^{-5}$ to 1 A/cm$^2$, and the third regime corresponds to higher currents.

To understand the interesting evolution of the distribution of the photon emission rate with the injection current shown in Fig. 2(c), we move on to the equation for the photon emission rate [29]:

$$R = \Phi \frac{1}{(c_n n)^{-1} + (c_p p)^{-1} + \Phi \tau_r} \quad (1)$$

Here, $\tau_r$ is the radiative lifetime of the excited state, $\Phi$ is the quantum efficiency, and $n$ and $p$ are the densities of electrons and holes in the vicinity of the color center. Both SiV and GeV centers have two charge states: neutral and negatively charged [15,21]. Therefore, as discussed above, the capture rate constants are roughly the same for both centers and equal $c_n = 1.7 \times 10^{-8}$ cm$^3$s$^{-1}$ and $c_p = 3.9 \times 10^{-7}$ cm$^3$s$^{-1}$ at room temperature [29]. Equation (1) shows that since the lifetime of the excited state for both SiV and GeV centers is in the nanosecond range [13,15], the SPEL rate $R$ is determined only by relatively slow electron and hole capture processes as shown below.

Figure 2(d,e) presents the simulated distributions of electrons and holes at different injection currents. In equilibrium, there are no free carriers in the i-region of the diode. As the bias increases, electrons are injected from the n-region and holes are injected from the p-region. Since $p_{eqp}$ is much higher than $n_{eqn}$, in the i-region, the density of holes is also higher than the density of electrons everywhere except the area near the i-n junction. Taking into account that the $c_p > c_n$, we obtain that at low injection levels, the optimum position of the color center, i.e., the position where the highest emission rate can be achieved, is in the very proximity of the i-n junction. As the current increases, the maximum photon emission rate $R_{max}$ that can be obtained at a fixed injection current also increases (see Fig. 3), and the optimum position of the color center is shifted to the n-region (see Fig. 2(c)), since more and more holes are injected into the i-region and these carriers penetrate deeper into the i-region.

When the density of holes at the i-n junction exceeds $\sim 3c_n n_{eqn}/c_p = 0.8\times 10^{10}$ cm$^{-3}$, the only limitation on the photon emission rate is the electron capture rate $c_n n_{eqn}$ (the doping limit) as follows from equation (1). Therefore, we observe a plateau in the input-output curve at current densities above $10^{-4}$ A/cm$^2$ (Fig. 3). The photon emission rate does not further increase with the pump current. The height of this plateau is mostly determined by the density $n_{eqn}$ of free electrons in the n-type region. The SPEL rate in the n-region is significantly higher than that in the i-region, and the optimum position of the color center appears exactly at the i-n junction [Fig. 2(c)]. In conventional semiconductors, such as silicon or gallium arsenide, we would observe this plateau even at extremely high bias voltages. However, in the considered diamond p-i-n diode with a 1-µm-thick i-region, the photon emission rate suddenly leaves the plateau and exceeds the doping limit at a current density of $J = 0.05$ A/cm$^2$ (Fig. 3).

The band diagram shown in Fig. 2(b) explains the unexpected rapid increase in the SPEL rate. At a bias voltage of $V = 4.5$ V (which corresponds to $J = 0.035$ A/cm$^2$), a hardly recognizable potential well for electrons is formed in the i-region in the vicinity of the i-n junction. Free electrons are accumulated in this potential well. The potential well for electrons is at the same time a potential barrier for holes. As the bias increases, the depth of the potential well increases and its bottom is shifted towards the p-region of the diode [Fig. 2(b)], which is clearly seen at biases of 4.7 and 4.8 V. The optimum position for the color center coincides with the bottom of the potential well, which is also shifted towards the p-region. Eventually, at very high bias voltages, the well is found in the very vicinity of the p-i junction.

The observed effect is similar to the effect of the gate electrode in a field effect transistor and could be referred as a self-gating effect. However, this is more an inverse self-gating effect rather than the direct one. In Fig. 2(b), one can see a flat-band p-region, while the band bending is strong in the i- and n-regions. The region of the potential well for electrons (potential barrier for holes) is a bridge between two regions with different types of conductivity (diffusion conductivity in the p-region and drift conductivity in the n-region) [39,40]. Due to the high asymmetry of the p-i-n structure, i.e., due to the huge difference in electrical properties of the n-type and p-type regions, the potential well is significantly deep at moderate and high bias voltages [40]. This gives the possibility to accumulate a very high density of free electrons in the well, namely one can inject nearly four orders of magnitude more electrons than the doping of the n-type region allows [Fig. 2(d)]. It is interesting that in this high-injection regime, the density of electrons in the well increases rapidly with the injection current and the slope $dR_{max}/dJ$ appears higher than at low injection levels (see the blue curve in Fig. 3). However, at $J = 48$ A/cm$^2$, $R_{max}$ reaches its maximum and then decreases. This roll-over looks very similar to that of semiconductor lasers [41], but the physical origin is entirely different. At a very high injection current, due to the relatively low resistivity of diamond, the bias voltage is very high, and a strong linear band bending is observed in both the n-type and p-type regions, which is very different from the well-known flat-band condition in conventional semiconductor diodes at very high injection levels [42]. In this case, the drift transport dominates over the diffusion one in the n-, i- and p-regions, and the potential well in the i-region gradually vanishes (see Fig. S3 in Supplemental Material [35]), which prevents electrons from accumulating in the i-region and eventually reduces the SPEL rate governed by the electron density (Fig. 3). Therefore, $R_{max}$ exhibits a maximum at $J = 48$ A/cm$^2$ ($V = 6.3$ V).

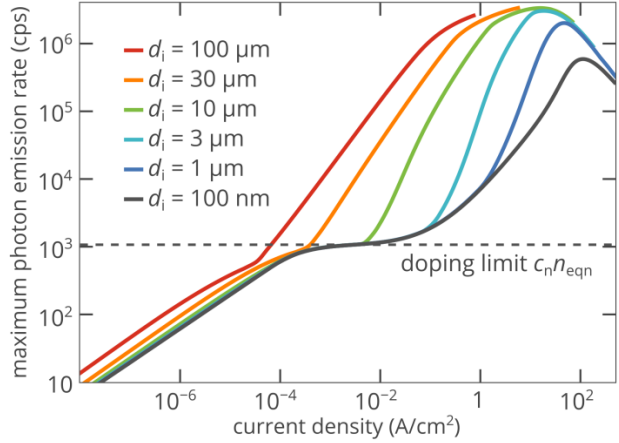

FIG. 3. Maximum photon emission rate $R_{max}$ that can be obtained from the color center in the p-i-n diode versus injection current for different thicknesses of the intrinsic region of the p-i-n diode. The quantum efficiency is assumed to be equal to 100%.

It is important to note that the superinjection effect can be efficiently exploited not only in the steady state but also in the pulsed mode. On-demand single photon sources should be triggered by short electrical pulses generating strictly one photon per pulse [43–45]. Therefore, the pulse length should be less than the 'recharge time' of the color center $((c_n n)^{-1}+(c_p p)^{-1}) \approx 1$ μs. However, the repetition rate can be limited by the response time of the diode. Our estimations of the capacitance of the diamond p-i-n diode with a 1-μm-thick i-region show that under superinjection conditions, it does not exceed 100 nF/cm$^2$ and is of the order of 1 nF/cm$^2$ at bias voltages above 4.9 V, which corresponds to $J > 15$ A/cm$^2$ in the steady state (see Fig. S4 in Supplemental Material [35]). Therefore, the response time of the diamond p-i-n diode is only of the order of 1 ns, which is nearly three orders of magnitude lower than the 'recharge time' of the color center, and thus it cannot limit the performance of the electrically driven single-photon source in the pulsed mode.

The strength of the self-gating effect greatly depends on the thickness $d_i$ of the intrinsic region of the p-i-n diode. For small thicknesses, such as $d_i = 100$ nm, the maximum rate is 3.4 times lower than at $d_i = 1000$ nm. Moreover, the maximum is achieved at a current density as high as 110 A/cm$^2$ (Fig. 3), which is not favorable for practical applications. On the contrary, as the thickness of the i-region increases, the depth of the potential well in the i-region increases. Accordingly, $R_{max}$ increases by 50% at $d_i = 10$ μm and is even higher at $d_i = 100$ μm (Fig. 3). More importantly, it is easier to achieve bright single-photon electroluminescence at large thicknesses $d_i$. For example, a current density of 19 A/cm$^2$ is needed to reach the photon emission rate of 1 Mcps at $d_i = 1$ μm. At $d_i = 10$ μm and 100 μm, this current reduces to 0.84 A/cm$^2$ and 0.09 A/cm$^2$, respectively. Despite that achieving a high emission rate at larger $d_i$ requires a higher bias voltage, the net power $JV$ supplied to the single-photon source appears to be lower, which is beneficial for practical applications. Our simulations show that at a photon emission rate of 1 Mcps, the single-photon emitting diode with a 100-μm-thick i-region consumes 210 times less power than the diode with a 1-μm-thick i-region (see Fig. S2 in Supplemental Material [35]). However, a very thick i-region acts as a series resistance, which leads to bias voltages greater than 10 V. Therefore, from a practical viewpoint, moderate thicknesses ~10-30 μm are probably more reasonable.

## III. DISCUSSION

The superinjection effect in diamond diodes can be very strong. However, to fully exploit its potential and achieve bright single-photon emission, it is extremely important to place the color center (using, for example, ion implantation) at the optimal position in the i-region of the p-i-n diode, since the position of the color center, which is a point defect in the crystal lattice, is fixed and cannot be tuned during diode exploitation. Figure 4(a) shows the simulated SPEL rate of the color center at five different positions in the i-region of the p-i-n diode with a 10-μm-thick intrinsic region. It is seen that at low injection currents, the highest brightness is achieved for color centers near the i-n junction ($z = 10$ μm). However, at current densities above 0.1 A/cm$^2$, the brightest color center is always found at a distance of less than 1 μm from the p-i junction, which corresponds to the bottom of the potential well for electrons discussed in the previous section. The maximum brightness of 3.3 Mcps (see the green curve in Fig. 3) is reached by the color center at a distance of about 400 nm from the p-i junction at a current density of 15 A/cm$^2$. Since it is easier to work with color centers near the diamond surface than with color centers located at a depth of 10 μm, we recommend to fabricate p-i-n diodes structures with the p-layer on top.

The above analysis clearly shows why bright SPEL could not be observed in the previous experimental studies of color centers in diamond p-i-n diodes. For example, SiV centers located at a distance of 0.3 μm from the n-i junction [21] are more than 500 times dimmer than at the optimal position near the p-i junction. According to our simulations, the maximum SPEL rate at this distance does not exceed 7000 cps, which, being multiplied by the quantum efficiency of 5% (which corresponds to the SiV center) and the collection efficiency of 5%, is only 18 cps at the photodetector. Thus, although the electroluminescence of SiV centers was demonstrated, the SPEL from a single SiV center has not been identified to date [20,21]. For comparison, a single SiV center located at a distance of 400 nm from the p-i junction can give 8400 cps at the photodetector, assuming the same quantum efficiency and collection efficiency. The range of positions of the SiV center and pump current densities at which bright single-photon electroluminescence can be achieved is shown in Fig. S5 in Supplemental Material [35].

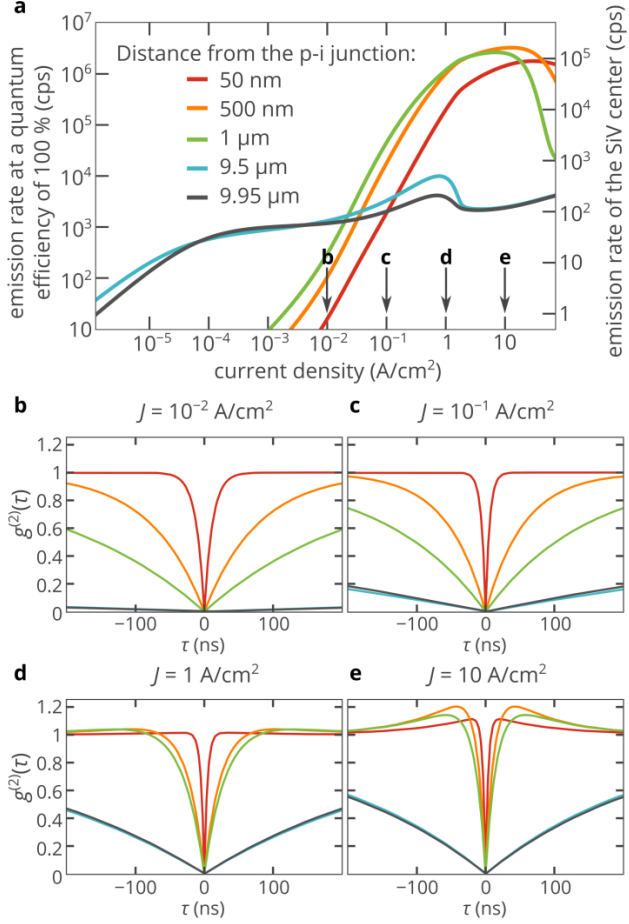

FIG. 4. (a) SPEL rate of the color center with two charge states (negative and neutral) and 100% quantum efficiency and SPEL rate of the SiV center (5% quantum efficiency) versus pump current simulated for five different positions of the color center in the p-i-n diode with a 10-um-thick i-region. (b-e) Simulated $g^{(2)}$ functions of the SiV center at four different injection currents. The notations are the same as in panel a. The lifetime of the excited state equals 1.2 ns and the lifetime of the shelving state is 100 ns [46].

We should note that the maximum SPEL rate is achieved at a current density of as high as $J_{opt} = 15$ A/cm$^2$ for the diode with a 10-μm-thick i-region (Fig. 3). At the same time, contacts to n-type diamond are known for their high specific contact resistance, which could suppress the superinjection. However, our numerical simulations show that the superinjection effect in the diamond p-i-n diode is very robust: the dependence of the SPEL rate on the current density is absolutely the same even if the n-type contact is Schottky with a very large Schottky barrier height (see Fig. S6 in Supplemental Material [35]). The contact resistance only increases the voltage drop across the diode by $J_n^{opt} \times R_{cont}(J_n^{opt})$ [47], where $J_n^{opt} = 0.4$ A/cm$^2$ is the electron current density at the n-type contact at the maximum SPEL rate and $R_{cont}(J_n^{opt})$ is the specific contact resistance of the n-type contact at this current. We note that in the considered diode structure, the electron current $J_n^{opt}$ is 38 times lower than the total current $J_{opt}$ due to the small thickness of the n-type layer and low density of free electrons in this layer. The contact resistance of the reverse biased Schottky contact is very high at low current densities. However, it steadily decreases as the electron current increases (see Fig. S7 in Supplemental Material [35]), and at an electron current density of 0.4 A/cm$^2$, $R_{cont}$ is as low as 200 $\Omega \cdot$cm$^2$ for a Schottky barrier height of as large as 2.5 eV. Thus, such a high barrier Schottky contact increases the voltage drop across the device by 80 V, which fully agrees with the self-consistent numerical simulations of the p-i-n Schottky diodes. Nevertheless, although the voltage drop increases, the electron and hole transport in the n-, i- and p-regions of the diode is unaffected, and the dependence of the SPEL rate on the current density is not altered. At the same time, we should emphasize that most ohmic contacts to n-type diamond provide much lower contact resistances at an electron current density of 0.4 A/cm$^2$. Moreover, recent experimental studies show that using a very thin diamond layer, which is $\delta$-doped with phosphorus [48] at a concentration of $\sim 10^{20}$ cm$^{-3}$ right below the metal contact, one can reduce the specific contact resistance to less than 0.1 $\Omega \cdot$cm$^2$ even at very low current densities [49,50], which completely eliminates the contact problem.

We also want to draw attention to the fact that the second-order autocorrelation function $g^{(2)}(\tau)$, which is very often measured in experiments, does not provide information about the brightness of the electrically pumped color center. The SPEL rate of the color center at different positions does not necessarily correlate with the characteristic time of the $g^{(2)}$ function, e.g., with the half-rise time $\tau_{1/2}$ defined as $g^{(2)}(\tau_{1/2}) = 1/2$. The reason for this is that the characteristic time $\tau_{1/2}$ is determined by the sum $(c_n n + c_p p)$ of the capture rates [30,36] and is, in fact, governed by the fastest process (usually, by the hole capture process). At the same time, the SPEL rate is determined by the slowest process (typically, by the electron capture process). This difference is especially pronounced for color centers near the p-i junction, where the density of holes is always higher than the density of electrons (see Fig. 2). In this region, the dynamics of single-photon emission is always faster than that in the center of the i-region or in the vicinity of the i-n junction regardless the injection level [Fig. 4(b-e)]. Moreover, we note that at every point of the i-region of the p-i-n diode, $c_p p / c_n n > 20$ for current densities above $10^{-4}$ A/cm$^2$. Therefore, the widely used $g^{(2)}$ function does not directly or indirectly show the photon rate of the single-photon emitting diode. However, the $g^{(2)}$ function can be efficiently used to measure the density of holes at the position of the color center and monitor the hole injection level.

## III. CONCLUSIONS

In summary, we have numerically demonstrated the superinjection effect in diamond p-i-n diodes. This effect gives the possibility to accumulate a high density of electrons, which is nearly four orders of magnitude above the equilibrium value in the n-type region of the diode, in the i-region in the vicinity of the p-i junction. Due to the exceptionally high activation energy of donors in diamond, the density of free electrons in the n-type region is 6–9 orders of magnitude lower than in most semiconductor devices. In addition, the density of holes in the p-type region is roughly 3-5 orders of magnitude higher than in the n-region. The high asymmetry between electrical properties of the n- and p-type regions combined with the low densities of carriers provides favorable conditions for the self-gating effect. Under high bias voltages, a potential well for electrons is formed in the i-region at a distance of about 200–500 nm from the p-i junction, which is similar to the effect of the gate in a field effect transistor. Electrons are accumulated in this potential well, which is beneficial for exploiting different 'local' effects, e.g., for developing single-photon emitting diodes based on color centers – point defects in the crystal lattice of diamond. Electroluminescence of the color center is based on the electron and hole capture processes. Since the SPEL rate is roughly proportional to the density of electrons in the vicinity of the color center, the superinjection effect can enhance the brightness of the electrically-driven sources by more than three orders of magnitude and reduce the power needed to drive the single-photon source, which is crucial for practical applications. Since it is extremely difficult to create a high density of free electrons in diamond at room temperature, the demonstrated effect can be used for the design of low-cost but efficient and bright single-photon devices for quantum information and quantum optics applications.

## ACKNOWLEDGEMENTS

The work is supported by the Russian Science Foundation (17-79-20421).

Supplemental Material for

# Superinjection in diamond p-i-n diodes: bright single-photon electroluminescence of color centers beyond the doping limit

Igor A. Khramtsov and Dmitry Yu. Fedyanin*

*Laboratory of Nanooptics and Plasmonics, Moscow Institute of Physics and Technology, 141700 Dolgoprudny, Russian Federation*

*E-mail: dmitry.fedyanin@phystech.edu

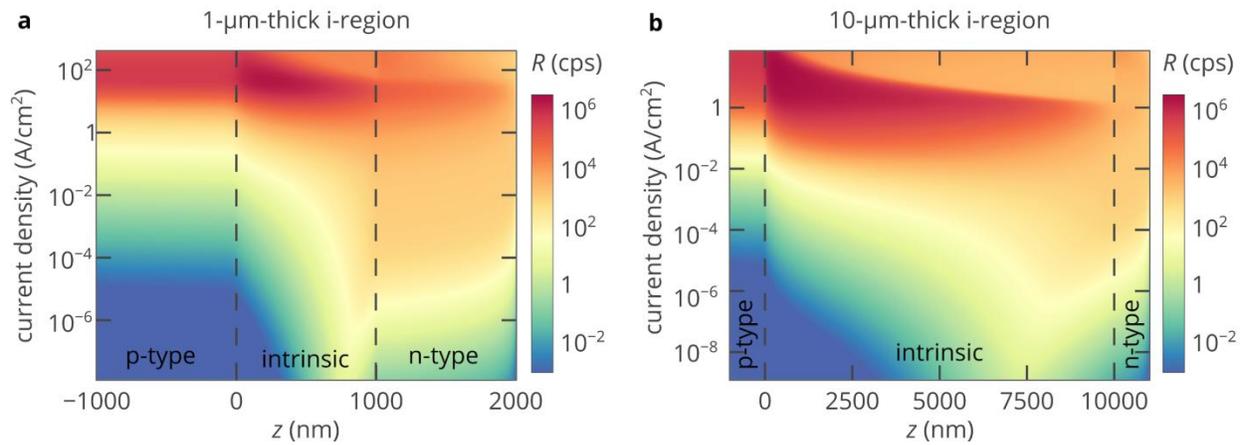

FIG. S1. Distribution of the single-photon emission rate $R$ as a function of the pump current for the diodes with 1-µm-thick (panel a) and 10-µm-thick (panel b) i-regions.

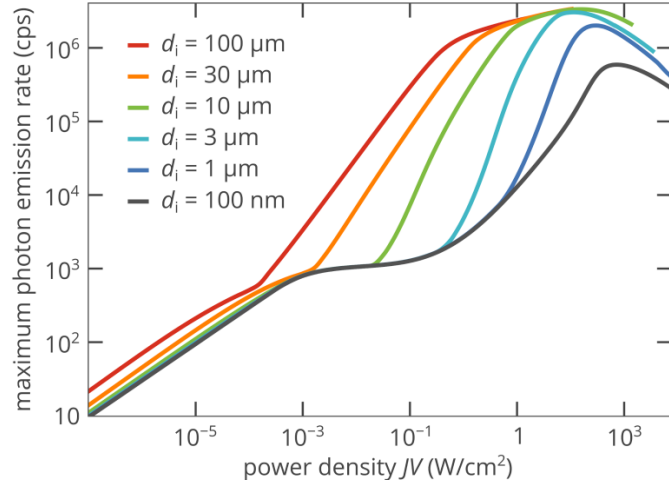

FIG. S2. Maximum photon emission rate $R_{max}$ that can be obtained from the color center in the p-i-n diode versus pump power density $JV$ for different thicknesses of the intrinsic region of the p-i-n diode. The quantum efficiency is assumed to be equal to 100%.

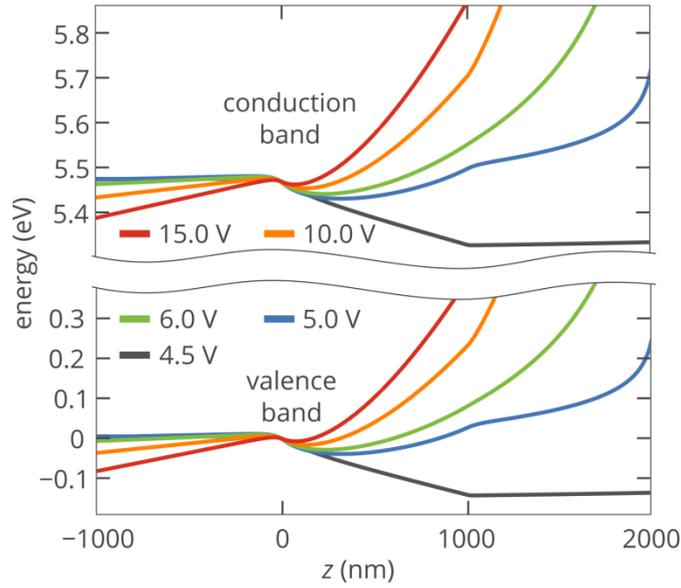

FIG. S3. Relative position of the energy band diagrams of the p-i-n diode with a 1-μm-thick i-region simulated at different bias voltages. For visual clarity, the energy band diagrams are aligned with each other by matching the conduction band edges at $z = -10$ nm.

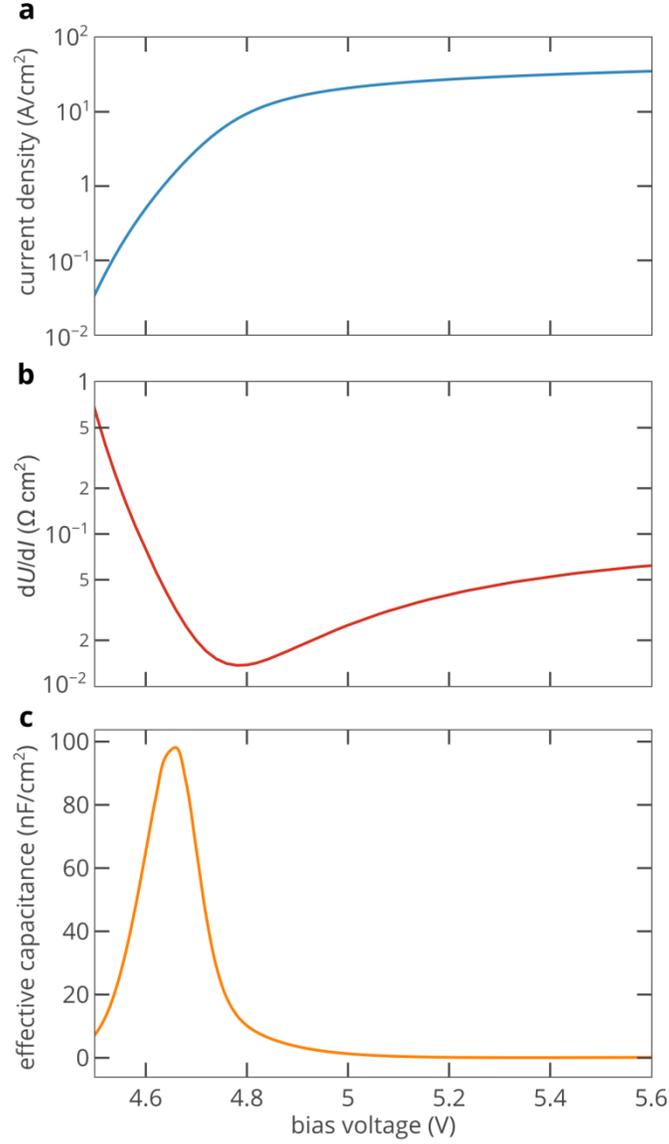

FIG. S4. (a) Current-voltage characteristics of the p-i-n diamond diode with a 1-µm-thick i-region. (b) Dependence of the differential resistance of the p-i-n diamond diode with a 1-µm-thick i-region on the bias voltage. (c) Effective capacitance per unit area as a function of the bias voltage applied to the p-i-n diode with a 1-µm-thick i-region. The effective capacitance $C_{eff}$ is calculated from the results of the numerical simulations using its thermodynamic definition $C_{eff} = 1/V \times dW/dV$, where $W$ is the energy stored in the p-i-n diode [51,52].

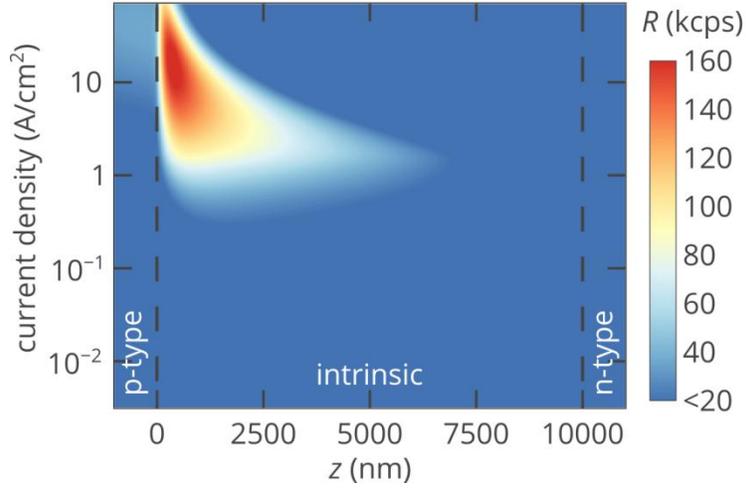

FIG. S5. Dependence of the SPEL rate of the SiV center on the color center position and the pump current density for the diamond p-i-n diode with a 10-µm-thick i-region.

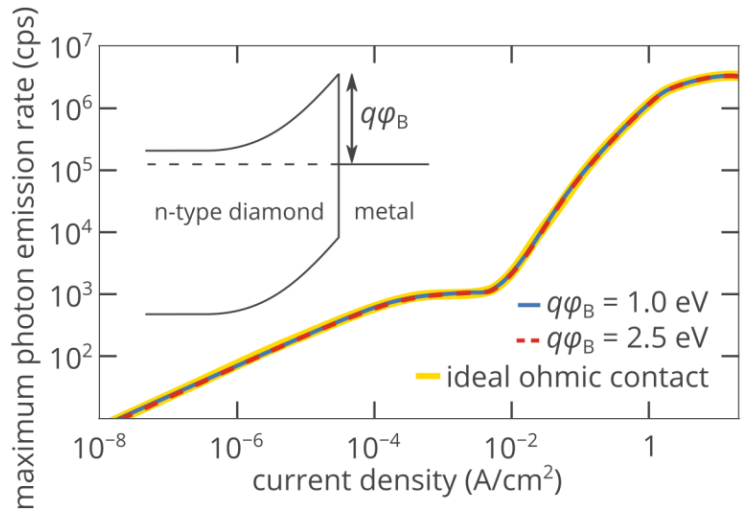

FIG. S6. Maximum photon emission rate $R_{max}$ that can be obtained from the color center versus injection current for the p-i-n diode with a 10-µm-thick i-region with three different contacts to the n-type diamond: yellow curve - ideal ohmic contact, blue curve - Schottky contact with a barrier height of 1.0 eV, and red dashed curve - Schottky contact with a barrier height of 2.5 eV. The diodes were simulated taking into account the field emission, thermionic-field emission, and thermionic emission electron transport through the Schottky contact [53-55].

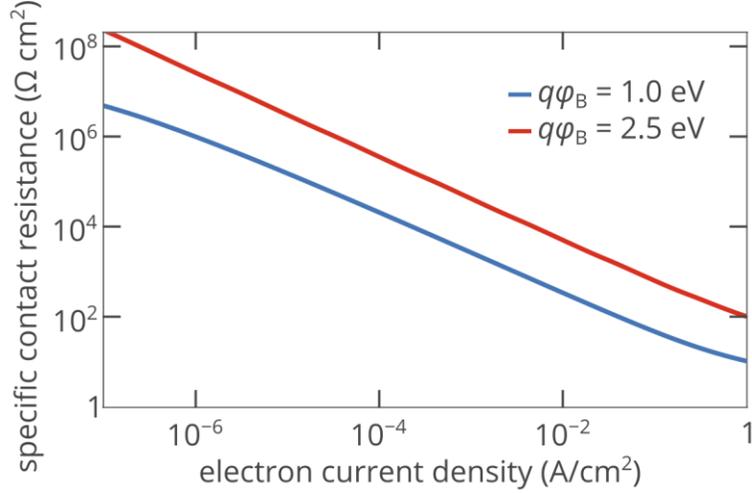

FIG. S7. Simulated dependence of the specific contact resistance of the reverse biased Schottky contact to n-type diamond on the electron current density for two Schottky barrier heights. The concentration of phosphorus in n-type diamond is equal to $10^{18}$ cm$^{-3}$, the donor compensation ratio equals $\eta_n = 10\%$. Other parameters of diamond used in the simulations are listed in Table S1. We note that the hole current density in the reverse biased n-type diamond/metal Schottky diode equals zero.

Table S1. Parameters used in the numerical simulations.

| Parameter | Value | Comment |
|---|---|---|
| Energy band gap of diamond | 5.47 eV | Ref. [56] |
| Dielectric constant of diamond | 5.7 | Ref. [56] |
| Density of donors in the n-type region | $10^{18}$ cm$^{-3}$ | |
| Activation energy of donors | 0.57 eV | Ref. [28] |
| Donor compensation ratio, $\eta_n$ | 10 % | |
| Density of acceptors in the p-type region | $10^{18}$ cm$^{-3}$ | |
| Activation energy of acceptors | 0.37 eV | Ref. [33] |
| Acceptor compensation ratio, $\eta_p$ | 1 % | |
| Electron mobility in the n-type and p-type regions | 740 cm$^2$/Vs | Refs. [56,57] |
| Hole mobility in the n-type and p-type regions | 660 cm$^2$/Vs | Refs. [56,57] |
| Electron mobility in the i-type region | 2500 cm$^2$/Vs | Ref. [37] |
| Hole mobility in the i-type region | 1200 cm$^2$/Vs | Ref. [37] |
| Electron capture rate constant, $c_n$ | 1.7×10$^{-8}$ cm$^3$s$^{-1}$ | Ref. [29] |
| Hole capture rate constant, $c_p$ | 3.9×10$^{-7}$ cm$^3$s$^{-1}$ | Ref. [29] |
| Density-of-states effective mass for electrons | 1.64$m_0$ | Ref. [58] |
| Heavy-hole effective mass | 0.67$m_0$ | Ref. [58] |
| Light-hole effective mass | 0.26$m_0$ | Ref. [58] |
| Split-off effective mass | 0.38$m_0$ | Ref. [58] |